\documentclass[aip,reprint]{revtex4-1}
\usepackage{graphicx}

\begin{document}

\title{A Distributed Magnetometer Network}

\author{John Scoville}
\affiliation{GeoCosmo Science Center, Mountain View, CA}
\affiliation{San Jose State University, Dept. of Physics, San Jose, CA, USA}
\affiliation{SETI Institute, Mountain View, CA, USA}
\affiliation{NASA Ames Research Center, Moffett Field, CA, USA}
\author{John Spritzer}
\affiliation{NASA Ames Research Center, Moffett Field, CA, USA}
\author{Friedemann Freund}
\affiliation{GeoCosmo Science Center, Mountain View, CA}
\affiliation{San Jose State University, Dept. of Physics, San Jose, CA, USA}
\affiliation{SETI Institute, Mountain View, CA, USA}
\affiliation{NASA Ames Research Center, Moffett Field, CA, USA}

\begin{abstract}
Various possiblities for a distributed magnetometer network are considered.  We discuss strategies such as croudsourcing smartphone magnetometer data, the use of trees as magnetometers, and performing interferometry using magnetometer arrays to synthesize the magnetometers into a large-scale low-frequency radio telescope.  Geophysical and other applications of such a network are discussed.
\end{abstract}

\maketitle

\section{Introduction}

A distributed network of magnetometers could be useful to many current and emerging areas of science.  One example that has partially motivated the present study is the observation of low-frequency pulses in the Earth's magnetic field\cite{Bleier2009}.  Experiments have shown that in some cases these pulses allow scientists to triangulate and forecast the epicenters of future earthquakes days or weeks in advance.  Currently, these pulses, which are able to penetrate the earth due to their extremely long wavelength, are observed by a limited number of relatively expensive and sensitive magnetometers whose maximum effective range is a few tens of km.  The broad spatial coverage offered by a distributed network could be key to observing these pulses.  Other magnetic anomalies also precede earthquakes - a month before the 3/11/2011 Tohoku earthquake in Japan, the reading from the magnetometer station closest to the epicenter drifted away from other nearby stations \cite{Takla2012}.  If a distributed 
network of low-cost or no-cost magnetometers was available to observe these signals, earthquake forecasting would become much more practical.  

In addition to earthquake forecasting, we note that there are many other potential applications and benefits.   In addition to observing transient and time-varying fluctuations in the magnetic field, the network could also identify spatial anomalies in the magnetic field.  These would, in turn, give clues as to the composition of nearby materials or assist in the discovery of natural resources which are often surveyed by measuring variations in electrical conductivity or by making a coarse measurement of the local magnetic field.

A distributed network potentially has a key performance advantage over highly sensitive single instruments: it could be used as a long-baseline interferometer.  An interferometery-based telescope offers the potential for a much higher resolution and sensitivity than traditional telescopes because it effectively has a diameter equal to the maximum separation of its components.  A distributed magnetometer network acting as an interferometer could potentially act as an omnidirectional radio telescope the size of the earth, or – if constellations of magnetometer-enabled satellites are put into orbit – perhaps even larger.  In addition to offering unprecedented performance to radio astronomers, the network could peer deep below the Earth's surface, performing tomography via low-frequency magnetic waves, like an MRI or CT-scan of the earth that offers a glimpse of structures and phenomena never before amenable to observervation by conventional magnetic geological surveys.

\section{Smartphones}

One intriguing possibility for a distributed magnetometer array is the crowdsourcing of magnetometer data from smartphones, most of which already include magnetometers.  These magnetometers are typically used to gather compass information.  This, along with the smartphone's built-in communication capability, makes a compelling case for using the devices to collect magnetic field data.  The smartphone app would be very simple – it would periodically send magnetometer and gps data to a central server-side repository, e.g. by using a web service.  Additionally, it could send the user alerts or other information from the server.  If the magnetometer data collection could be incorporated as part of an app that already has a large base of users, the data could collected immediately, and at practically no cost.

The initial database/baseline values could be based on calculating the geomagnetic field for the location of interest and (if the data exists) add to a magnetic anomaly map.  The measurement would then use the mapped intensity as a baseline value. 

If, say, interferometry is being performed, the processing required at the central data repository might be significantly greater.  However, a basic system could be implemented without a great deal of complexity.  The prototype network might not start with a fully crowdsourcing-ready implementation but instead start with an array of stationary magnetometers.  An array of Hall-effect might be a good test for the algorithms that would eventually combine measurements from smartphone magnetometer data, since a distributed magnetometer network could theoretically provide improved spatial resolution in addition to enhanced sensitivity.

The only downside to using cellphones for an array is that the sensitivity of these magnetometers usually isn't great - 150-300nT is typical.  However, the sensitivity may improve with subsequent generations of phones.  If there was a demand for more sensitive magnetometers, hardware manufacturers could readily incorporate them into the devices - magnetometer chips with sensitivity in the nanotesla range are currently available for around 20 dollars.

To observe a wide variety of transient features in the local magnetic field, the array would ideally have nanotesla sensitivity and as high a sampling rate as possible, probably in the 100Hz to 10kHz range.  In addition to resolving relatively fast transient signals, a higher sampling rate could translate into higher effective sensitivity.  

Ideally, the network would resolve magnetic field vectors, but in many cases the total magnetic field strength is sufficient.  The total magnetic field has the advantage of being rotationally invariant such that the angular orientation of the device does not affect the measurement.  If the phone's angular orientation is tracked via its internal gyroscope, then 3-axis measurements should be possible.  

Accurately keeping track of the approximate location of thousands of smartphones via GPS is relatively simple, so long as the phones have good reception.  One company, indooratlas.com, is already crowdsourcing the mapping of local magnetic fields for the purpose of assisting navigation indoors, where GPS is less effective.  One can always use the phone's gps and accelerometer data to tell whether the phone is moving or stationary, and, if motion proves problematic, use this information to collect data from devices that are stationary.

To conserve bandwidth, the device doesn't have to constantly transmit as the samples are being taken.  The data can be accumulated, compressed, and sent in batches as necessary.  Once the angular orientation of the phone has been corrected for, the magnetic field measurements should not exhibit a huge degree of fluctuation, so much of the data will be redundant and highly compressible.  

\section{Trees}

Surprisingly, trees can act as extremely sensitive magnetometer antennae.  Anthony Fraser-Smith of Stanford University demonstrated in 1977 that magnetic field fluctuations could be recorded by simply inserting a pair of electrodes into a oak tree\cite{FraserSmithTrees} and measuring the potential difference across the electrodes.  Astoundingly, the tree was found to have better sensitivity to many features of the magnetic field than a high-end magnetometer.  The tree was used to record PC1 pulsations from the geomagnetic field with great accuracy and temporal response.  Since highly sensitive magnetometers are delicate instruments that frequently cost tens of thousands of dollars each, the cost savings that could be achieved by using trees could be significant.

Trees are potentially an even cheaper and more sensitive solution than smartphone magnetometers, even though the instrumentation needed to access them is significantly more involved.  Three weeks prior to the Loma Prieta earthquake, a tree-based magnetometer recorded variations in the local magnetic field that were 32 standard deviations above its baseline.  A traditional magnetometer recorded variations two weeks prior.

Several obstacles would need to be overcome before trees could be reliably used.  First, the trees would almost certainly need to be calibrated since there would surely be sizable variations in the voltages produced by different trees in response to magnetic fields.  Secondly, more work needs to be done to determine whether the tree is measuring only the magnetic field, or some combination of magnetic and electric fields in conjunction with its local environment, water, nutrients, etc.  

Trees lack the convenient communication capabilities that are built into phones, and would likely require a preamplifier for optimal performance.  One means of overcoming this obstacle is to use a smartphone in conjunction with the tree potential, recording the voltage fluctuations as an audio signal, using a small amplifier if necessary, and using the smartphone's communications capability to relay the data to the server.  Obviously, this would require the use of a dedicated smartphone, but the tree antenna could potentially achieve higher levels of sensitivity.  

Yet another possibility would be to use the tree's voltage, amplified if necessary, to directly power an LED.  The amount of light output from the LED would then indicate the voltage across the tree's electrodes.  A large number of trees in an area could be wired up with LEDs and simultaneously recorded by a stationary video camera connected to a computer.  A computer, running an application that monitors the video pixels corresponding to LEDs, creates a timeseries of numerical values indicating the level of light emitted by each LED.  This timeseries, appropriately normalized, could provide magnetometer data.  One downside of this approach, aside from the adverse effects of fog, dust, etc. - is that its sensitivity is limited by the dynamic range of the camera – typically 8 or 16 bit – as well as the camera's frame rate, conventionally 30Hz.  Also, the dynamic range of the LED would have to match the CCD being used.  If the details could be worked out, this could be an inexpensive way to obtain large 
amounts of data.

\section{The Fast Fourier Transform Telescope and Interferometry}

Once magnetic field data is collected and communicated back to the central repository, it must be synthesized into a cohesive map of the magnetic field at the surface of the earth.  There are many compelling reasons to carefully synthesize the data rather than simply patching together a large number of individual measurement taken at different locations.  Combining measurements allows greater sensitivity.  So long as noise and measurement error across different magnetometers are not correlated, simple statistics dictate that the net accuracy should steadily improve as more data points are collected.

Another compelling reason is that, by combining a large number of magnetometers, we can move beyond mere point measurements of magnetism and enable a form of tomography that will allow us to actually see an image of the magnetic field, akin to what one might see using a radio telescope, or, at the frequencies of visible light, our eyes.  This could potentially reveal much more than a simple point measurement, for instance, by imaging features of the Earth's interior that are currently completely inaccessible due to their depth.  In the case of low-frequency pre-seismic magnetic pulses such tomography could offer more information than simply triangulating the source of the pulses.

Both of these objectives may be realized by implementing a so-called “Fast Fourier Transform Telescope” (FFTT) or “omniscope” \cite{FFTT}.  This is an all-digital telescope that combines the key advantages of both large single-dish radio telescopes and distributed interferometers.  As Hippolyte Fizeau had already observed by 1868, the lenses and mirrors of telescopes are simply performing a physical analog to a Fourier transform.  This marked the beginning of astronomical interferometry.  A Japanese team in 1993 used an all-digital technique to take 8x8 pixel images of the Crab nebula with a 2-D 8x8 Spatial Fourier Transform processor.

Tegmark and Zaldarriaga propose that this design should be used for large-scale telescopes for a number of reasons.  One, it is completely digital and does not rely on any lenses, mirrors, or dishes.  Furthermore, unlike traditional interferometers, whose cost becomes dominated by the computing power needed to calculate $N^2$ pairwise correlations between individual elements, if the observation points are on a rectangular grid, this cross-correlation may be performed using the FFT in an amount of time proportional to $N log N$.  As such, the design facilitates the construction of interferometers that include a number of elements that would otherwise be impractical.  In the case of observation points that are not on a rectangular grid, a non-uniform fast fourier transform could serve the same function, although this could incur additional computational expense.

The only obstacle to this sort of design in the past has been a lack of computer power to perform the FFT, but this is no longer the case, and such instruments offer broad new horizons.  Compared to traditional interferometers, an FFTT has the advantage of being able to digitally focus on a very narrow beam while being much less expensive due to its reduced computational footprint.  Compared to large single telescopes (or, in our case, single magnetometers) an FFTT offers vastly superior sensitivity at a much lower cost for large-area, omnidirectional surveys.    This makes the FFTT an ideal means of synthesizing data from a distributed magnetometer network into highly accurate magnetic tomography.

\section{Conclusions}

Recent advances in computing power and consumer technology have created an environment in which the implementation of a very large scale distributed magnetometer network has become practical.  Such a system could allow observations that are currently completely infeasible.

Although we have focused in this discussion on applications related to geophysical tomography, many other applications exist.  Astrophysics, the first discipline to use the FFTT, could benefit greatly from the application of interferometry to magnetometers - no telescope has ever been constructed that can image such low-frequency electromagnetic waves.  Such a device would reveal regions of the spectrum that have not previously been imaged, possibly leading to the discovery of new features of the universe.

In addition to imaging, the network could also be useful in tracking magnetically susceptible objects, e.g. items made out of metal.  The movement of planes, cars, and other objects could be inferred from magnetometer data.  The magnetic flux near a highway, for instance, would provide a means of measuring the flow of traffic.

\end{document}